\documentclass[preprint,journal]{vgtc}                
\ifpdf
  \pdfoutput=1\relax                   
  \usepackage{graphicx}                
  \DeclareGraphicsExtensions{.pdf,.png,.jpg,.jpeg} 
\else
  \usepackage{graphicx}                
  \DeclareGraphicsExtensions{.eps}     
\fi%

\graphicspath{{figures/}{pictures/}{images/}{./}} 

\usepackage{microtype}                 
\PassOptionsToPackage{warn}{textcomp}  
\usepackage{textcomp}                  
\usepackage{mathptmx}                  
\usepackage{times}                     
\usepackage{cite}                      
\usepackage{tabu}                      
\usepackage{booktabs}                  

\usepackage{xcolor}
\usepackage{paralist}
\usepackage{amssymb}
\usepackage{float}
\usepackage{placeins}
\usepackage{scrextend}

\DeclareMathAlphabet{\mathcal}{OMS}{cmsy}{m}{n}

\usepackage[draft=true]{minted}
\usepackage[hyphens]{url}
\usepackage[hidelinks]{hyperref}
\hypersetup{breaklinks=true}

\usepackage{multirow}

\newenvironment{rev}[0]{%
    \leavevmode\color{black}\ignorespaces
}{}

\newcommand{\name}{MultiVision}

\usepackage{xspace,xpunctuate}
\newcommand{\ie}{{i.e.,}\xspace}
\newcommand{\eg}{{e.g.,}\xspace}
\newcommand{\ea}{{et~al\xperiod}\xspace}

\newcommand{\tableField}{Table Field}
\newcommand{\mvRec}{Multiple-View Recommender}
\newcommand{\controlP}{Control Panel}
\newcommand{\chartEd}{Chart Editor}
\newcommand{\chartIdea}{Chart Ideas}
\newcommand{\mvView}{Dashboard View}
\newcommand{\crossChart}{Cross-Chart Interactions}

\usepackage{amsmath}
\DeclareMathOperator{\sign}{sign}

\usepackage{tikz}
\definecolor{nodeBG}{RGB}{29,40,97}
\newcommand*\circled[1]{\tikz[baseline=(char.base)]{
            \node[shape=circle,draw, fill=nodeBG,inner sep=0pt] (char) {\textcolor{white}{#1}};}}

\usepackage{amsmath}

\usepackage{array}

\usepackage[ruled,vlined]{algorithm2e}
\SetKwComment{Comment}{$\triangleright$\ }{}

\onlineid{1035}

\vgtccategory{Research}
\vgtcpapertype{System (Area 5: Data Transformations)}

\title{\name: Designing Analytical Dashboards \\ with Deep Learning Based Recommendation}

\author{
Aoyu Wu, Yun Wang, Mengyu Zhou, Xinyi He, Haidong Zhang, Huamin Qu, and Dongmei Zhang
}
\authorfooter{
\item
 A. Wu and H. Qu are from Hong Kong University of Science and Technology. E-mail: \{awuac, huamin\}@cse.ust.hk. This work was conducted when A. Wu was an intern at Microsoft Research Area.
\item
 Y. Wang, M. Zhou, X. He, H. Zhang, and D. Zhang are from Microsoft Research Area. Email: \{wangyun, mezho, v-hexin, haidong.zhang, dongmeiz\}@microsoft.com. Y. Wang is the corresponding author.
}

\shortauthortitle{Biv \MakeLowercase{\textit{et al.}}: Global Illumination for Fun and Profit}

\abstract{
We contribute a deep-learning-based method that assists in designing analytical dashboards for analyzing a data table.
Given a data table,
data workers usually need to experience a tedious and time-consuming process to select meaningful combinations of data columns for creating charts.
This process is further complicated by the needs of creating dashboards composed of multiple views that unveil different perspectives of data.
Existing automated approaches for recommending multiple-view visualizations mainly build on manually crafted design rules,
producing sub-optimal or irrelevant suggestions.
To address this gap,
we present a deep learning approach for selecting data columns and recommending multiple charts.
More importantly,
we integrate the deep learning models into a mixed-initiative system.
Our model could make recommendations given optional user-input selections of data columns. 
The model, in turn, learns from provenance data of authoring logs in an offline manner.
We compare our deep learning model with existing methods for visualization recommendation and conduct a user study to evaluate the usefulness of the system.
} 


\CCScatlist{ 
 \CCScat{K.6.1}{Management of Computing and Information Systems}%
{Project and People Management}{Life Cycle};
 \CCScat{K.7.m}{The Computing Profession}{Miscellaneous}{Ethics}
}

\teaser{
    \centering
    \includegraphics[width=1\linewidth]{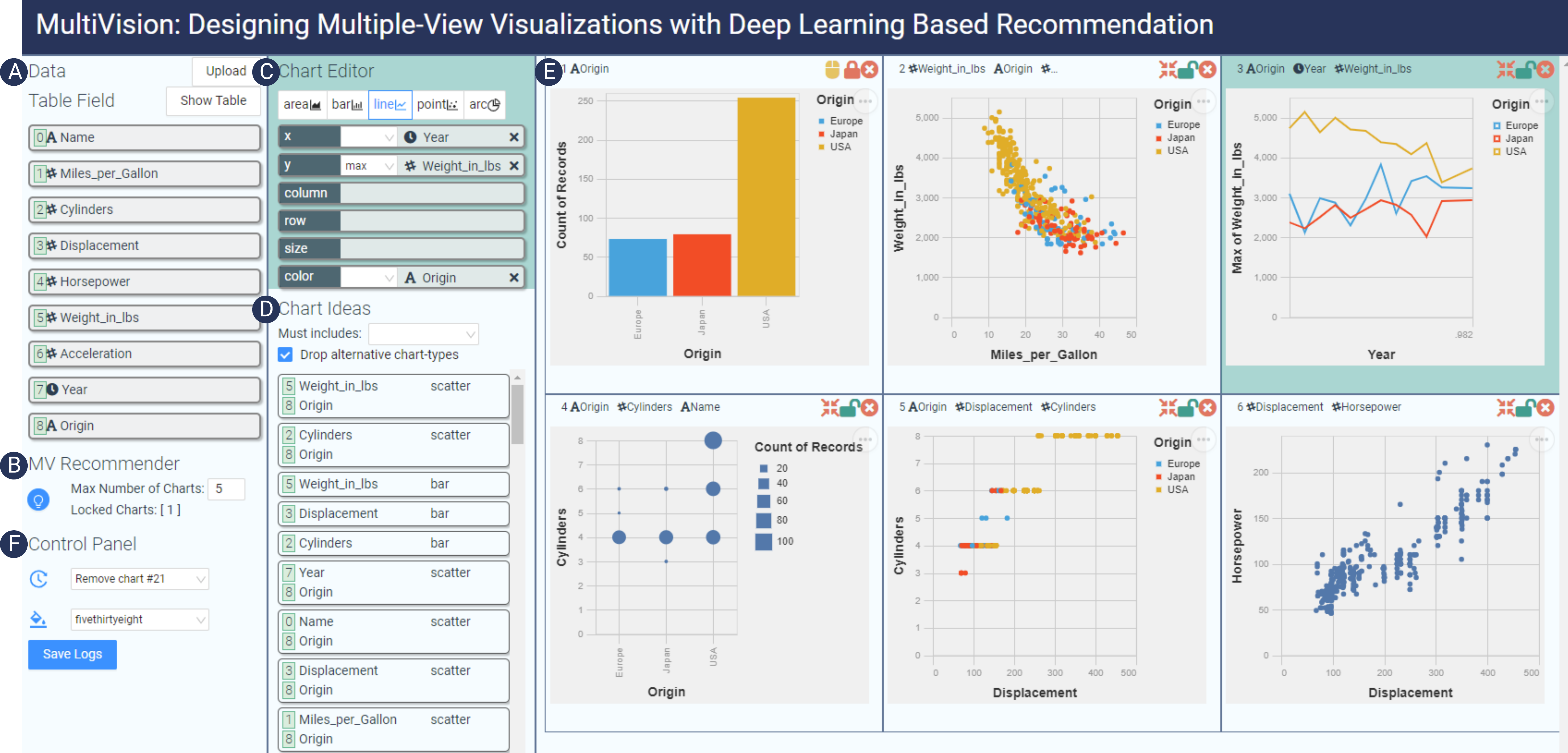}
    \caption{The interface: 
    {\protect\circled{A}} When uploading a tabular dataset, the \tableField~shows data columns with the corresponding data type;
    {\protect\circled{B}} Clicking the \mvRec~will generate a dashboard containing a user-specified number of charts. The recommendation is conditioned on, if any, user-locked charts;
    {\protect\circled{C}} Users could specify the chart type, encoding channels, and data transformation in the \chartEd;
    {\protect\circled{D}} \chartIdea~recommends charts based on the current dashboard;
    {\protect\circled{E}} The \mvView~displays the charts in an interactive grid layout;
    and {\protect\circled{F}} Users could restore the history, change the theme, and save logs in \controlP.}
    \label{fig:interface}
}

\vgtcinsertpkg

\begin{document}
\firstsection{Introduction}
\maketitle
Data visualization is becoming an increasingly used technique for analyzing a dataset,
contributing to the creation of visualization tools that empower laypeople and democratize data analytics.
In those visualization tools,
the basic workflow typically starts with uploading a dataset in a tabular format,
followed by selecting the data of interests and creating corresponding charts (\eg Excel~\cite{excel} and Tableau~\cite{tableau}).
However, users usually need to engage in a tedious and time-consuming process to explore different data selections through trial and error.
Besides,
it requires experience and expertise to create visualizations that effectively facilitate data analysis~\cite{qin2019making}.
This process is further complicated by the demands for creating multiple-view visualizations that enable simultaneous exploration of the same data from different perspectives~\cite{roberts2007state}.
Due to those challenges,
there have been huge research efforts in developing visualization recommendation systems that assist laypeople in conducting visual data exploration.

Most existing recommendation systems build upon visualization design knowledge gained from empirical experiments.
For instance,
APT~\cite{mackinlay1986automating}, Show Me~\cite{mackinlay2007show}, and Voyager~\cite{wongsuphasawat2015voyager} recommend visual encodings according to their effectiveness rankings based on perceptual principles. 
Recent systems extend those approaches with data recommendation support, \ie to select data columns to be visualized.
Those systems (\eg Voder~\cite{srinivasan2018augmenting}, DataShot~\cite{wang2019datashot}, Calliope~\cite{shi2020calliope}) associate charts with insights, thereby proposing hand-crafted metrics to rank and recommend insights.
However,
hand-crafted metrics face limitations such as sub-optimal performances and high requirements for domain expertise~\cite{wu2021survey}.

In response, 
recent research starts to propose machine learning (ML) methods that learn to recommend visualizations from large-scale corpora.
Existing approaches such as Draco~\cite{moritz2018formalizing}, Data2Vis~\cite{dibia2019data2vis}, and VizML~\cite{hu2019vizml} mainly focus on recommending visual encodings of a single chart.
Although proven useful, they do not support data recommendation.
This limitation results in a quasi-automated process that users need to manually select data columns prior to automated recommendation.
However, manual selection is time-consuming and challenging due to two reasons:
First, not every data column counts. 
The number of possible charts grows exponentially as the cardinality of data columns increases. 
Therefore, it is necessary to select data columns that are more ``worthy of'' visual analysis.
Second, one chart does not fit all. 
It is crucial to design dashboards containing multiple charts that satisfy different criteria such as diversity and consistency~\cite{qu2017keeping}. 
Some combinations of charts might not be interesting or meaningful~\cite{wang2000guidelines}. 


To address the above challenges,
we aim to propose deep-learning-based approaches for recommending an analytical dashboard.
\begin{rev}
Dashboards are typically a multiple-view visualization (MV),
and we interchangeably use those two terms throughout this paper.\end{rev}
We focus on the selection of MVs,
\ie to select important and meaningful data columns.
Different from existing ML-based recommenders that generate charts in an end-to-end manner and hardly enable user control,
we take a mixed-initiative perspective that integrates automated recommendation into an interactive system for authoring dashboards and exploring a dataset.
In doing so, we hope to leverage the combined power of human agency and machine automation to solve the challenging task of constructing effective dashboards.



We present \name{}, a mixed-initiative system for designing multiple-view visualizations for analyzing tabular datasets.
\name{} features deep-learning models for assessing the qualities of charts,
whereby recommending best-scored charts.
Specifically,
we develop two deep-learning models for assessing a single chart and assessing multiple charts, respectively.
The first model builds on a large corpus consisting of more than 200k Excel data tables and charts~\cite{zhou2020table2charts}.
Due to the lack of large-scale datasets of MVs,
we design the second model for learning from provenance data, \ie authoring logs in \name.
More importantly,
we design and implement an interactive system for users to interact with automated recommendations (\autoref{fig:interface}).
Once users edit charts in the MV,
the system automatically makes recommendations based on the current MV to facilitate data exploration.
\begin{rev}
While the deep-learning models focus on recommending data column selections,
the system interface allows users to adjust the presentation of MVs (\ie visual encodings and layouts) and interact with the MV (\eg cross-chart filtering and highlighting).
\end{rev}

To evaluate~\name,
we conduct an experiment to compare our model with baseline approaches in existing ML-based visualization recommenders.
We further conduct a user study with 12 participants to demonstrate the usefulness of our system,
whereby discussing implications for future research regarding the challenging and important problem of recommending and authoring dashboards.
Our codes are open-sourced\footnote{\url{https://github.com/Franches/MultiVision}}.
In conclusion, the contributions of this work include:
\begin{compactitem}
    \item A deep-learning-based approach for recommending multiple-view visualizations
    \item A mixed-initiative system that integrates automated recommendation into interactive authoring of multiple-view visualizations and data exploration
    \item An experiment and a user study that demonstrates the effectiveness of our deep learning models and system, respectively
\end{compactitem}


\section{Related Work}
This work is related to visualization recommendation, dashboards, visualization authoring tools, as well as mixed-initiative systems.

\subsection{Visualization Recommendation}
Decades of research have proposed many visualization recommendation systems that are thoroughly discussed in recent surveys~\cite{qin2019making,zhu2020survey,wu2021survey}.
There are growing research interests in leveraging machine learning (ML) for visualization recommendation~\cite{wang2020applying}.
ML-based methods learn visualization design knowledge from large-scale corpora,
surpassing traditional rule-based approaches and even laypersons in some cases~\cite{hu2019vizml}.
For instance,
Data2Vis~\cite{dibia2019data2vis} and VizML~\cite{hu2019vizml} formulate a prediction problem,
\ie to predict the visual encodings given data columns to be visualized.
These prediction-based methods output a single chart,
which can be insufficient in unveiling different perspectives of data.

In contrast, other approaches learn to assess the ``goodness'' of visualizations,
whereby recommending top-k assessed charts.
DeepEye~\cite{luo2018deepeye} trains a binary classifier to determine whether a chart is good or bad and subsequently ranks charts by pair-wise comparisons.
Instead of rankings,
other methods output scores that better reflect the absolute qualities of charts.
Draco~\cite{moritz2018formalizing} uses RankSVM to learn the weighted hand-crafted constraints to obtain an overall score of visualization specifications.
LQ2~\cite{wu2021learning} proposes a Siamese neural network to predict chart layout qualities from crowdsourcing comparison data.
\begin{rev}
Those approaches recommend visual encodings,
where the output has fixed cardinality.
We address a different problem, \ie to recommend data column selections and chart combinations,
which has varying cardinality,
\eg charts might encode a varying number of data columns.
Our method builds upon similar Siamese network structure,
but extends it with recurrent neural networks and new chart embedding modules to overcame the above challenges,
outperforming the naive Siamese structure in our experiment.
\end{rev}


\subsection{Dashboards and Multiple-View Visualizations}
A multiple-view visualization (MV) composes multiple visualizations into a single cohesive representation that facilitates analyzing different perspectives of data~\cite{roberts1998encouraging}.
One common genre of MVs is dashboards,
which is a common technique for visual analytics~\cite{sarikaya2018we}.
Despite their prevalence,
there exist few guidelines on designing effective MVs or dashboards.
Baldonado~\ea~\cite{roberts1998encouraging} drew upon workshop discussions to present eight guidelines for using multiple views in information visualizations. 
Qu~\ea~\cite{qu2017keeping} emphasized the consistency constraints among multiple charts.
Besides,
several work studies how to extend MV to multiple devices~\cite{langner2017v,sadana2016designing} or large displays~\cite{langner2018multiple}.
However,
those studies investigate design guidelines as qualitative reflections from empirical experiments rather than quantitative metrics.
It remains unclear how to quantitatively encode those guidelines to promote automated design.

In response,
recent research starts to quantify the MV and dashboard designs from a data-driven perspective.
Al-maneea and Roberts~\cite{al2019towards} quantified the layout designs as seen from visualization publications.
Chen~\ea~\cite{chen2020composition} investigated the composition and configuration patterns of MV,
whereby developing a recommendation system for suggesting an exemplar design from visualization publications.
LADV~\cite{ma2020ladv} generates dashboard visualizations by recognizing chart types from an input image or sketch.
While helpful, those approaches mainly focus on the \textit{presentation} of views such as layout design or colour palette.
Different from them,
we study \textit{selection} of views,
that is,
how to select multiple charts from many candidates given a data table in order to facilitate data analysis.
Inspired by Draco~\cite{moritz2018formalizing},
we seek to propose quantitative metrics to encode MV design guidelines and learn to weight different metrics to recommend top-k MV designs.

\subsection{Visualization Authoring Tools}
Researchers have investigated ways of making it easy to design and create data visualizations. 
Numerous efforts support visualization design in the form of interactive authoring systems.
For example, early research such as 
SageBrush \cite{roth1994interactive} let users create visualizations by choosing chart types through drag-and-drop operations. 
Recently, more advanced tools such as Lyra~\cite{satyanarayan2014lyra}, iVisDesigner~\cite{6876042}, DDG~\cite{Kim_2017_7536218}, InfoNice~\cite{Wang:2018:IEC:3173574.3173909}, Data Illustrator~\cite{Liu:2018:DIA:3173574.3173697}, Charticulator~\cite{Ren_2019_8440827} provide users with power of more customized visualization design of encodings and styles. 

\begin{figure*}[!t]
	\centering
	\includegraphics[width=1\linewidth]{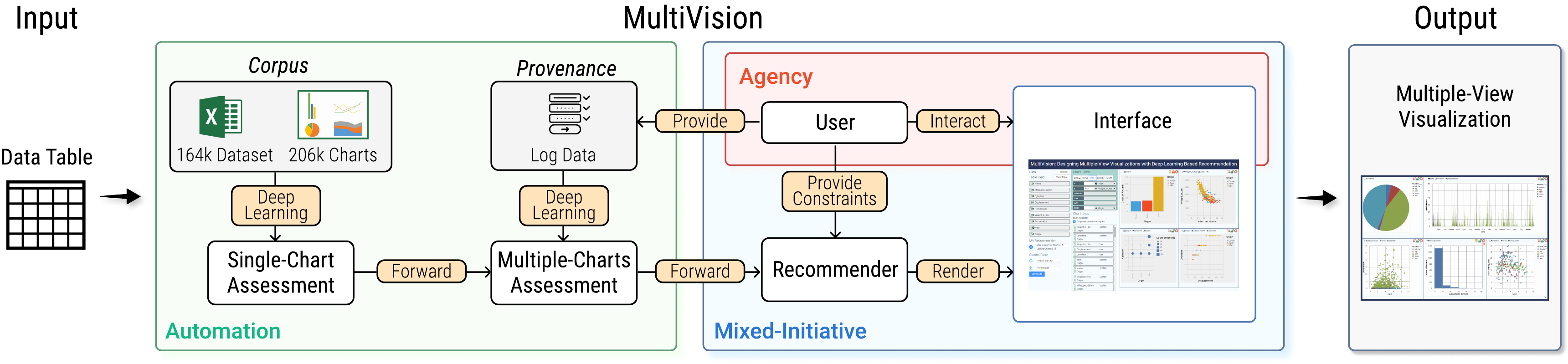}
	\caption{\name~is a mixed-initiative system that generates a multiple-view visualization given a data table.
	The automation consists of deep learning models that learn to assess the quality of a single chart and multiple charts from a chart corpus and provenance data.
	The assessment models, combined with user-input constraints, make recommendations of MVs, which are then rendered in the interface.
	The agency can interact with the recommended results in the interface,
	and the editing logs, in turn, are fed into the deep learning models in an offline manner.}
	\label{fig:overview}
\end{figure*}

However, those tools mainly focus on encodings and designs of a single visualization.
\begin{rev}
Tulip~\cite{auber2004tulip} and Cytoscape~\cite{franz2016cytoscape} support authoring MVs for graph data, which are incapable of tabular datasets.\end{rev}
Commercial tools like Tableau and Microsoft Power BI have enabled users to encode data with different visual forms, which are further arranged together into MVs such as dashboards. 
Similarly, we enable users to create MVs for visual analytics with the additional power of ML-based recommendations.

In the field of data storytelling, researchers have investigated approaches of composing multiple visualizations in a logical form. 
For example, Hullman et al. \cite{hullman2013deeper} and Kim et al. \cite{kim2017graphscape} have proposed algorithms to generate a sequence of visualizations to support storytelling and sequencing. 
DataShot~\cite{wang2019datashot} and Calliope~\cite{shi2020calliope} organize visualizations into topics based on data insights. 
Different from them, we aim to assist users in designing MVs for data analysis.

\subsection{Mixed-Initiative System for Visualizations}
Mixed-initiative systems allow the human agency to collaborate with computer automation~\cite{horvitz1999principles}.
In the field of data visualization,
many systems have been proposed to facilitate the design and use of visualizations.
For example,
VizAssist~\cite{bouali2016vizassist} guides users to find a relevant visualization through an interactive genetic algorithm that adapts to user needs.
\begin{rev}
Bylinskii~\ea~\cite{bylinskii2017learning} proposed a method for predicting and showing the visual importance of visualizations during interactive authoring.\end{rev}
Other applications include managing ambiguity in natural language~\cite{gao2015datatone} or extracting data from chart images~\cite{jung2017chartsense}.

\name{} differs from the above systems in two aspects.
Firstly,
human input or intervention is made compulsory in those systems.
In contrast,
\name{} provides more significant value-added automation where human input is not mandatory.
Instead,
users can optionally request recommendations based on their current MV to obtain customized results.
More importantly,
we propose two recommendation strategies,
including the passive recommendation that is triggered upon request and the active recommendation that automatically updates with user changes.
We draw on the participants' feedback in the user study to discuss our lessons learnt.

\section{Overview}

\begin{figure}[!t]
	\centering
	\includegraphics[width=1\linewidth]{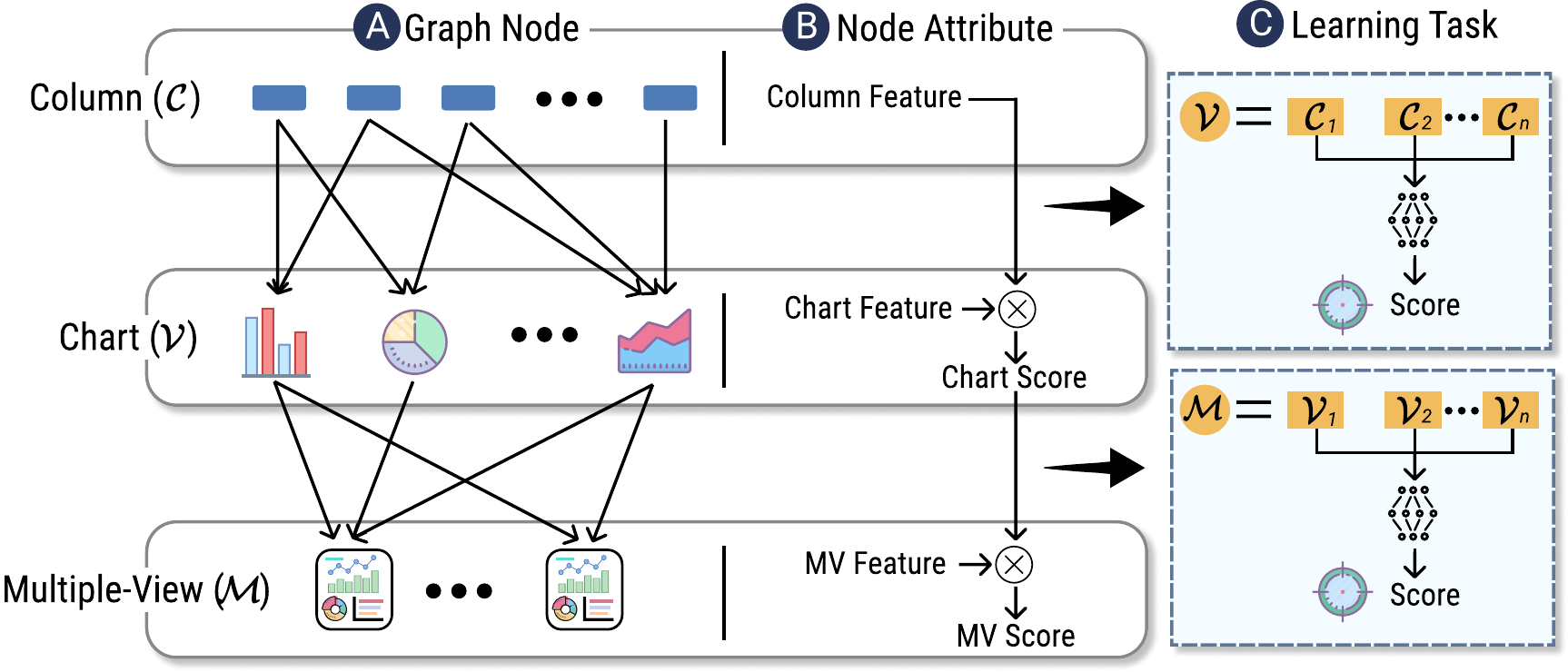}
	\caption{
	We abstract the recommendation problem into machine learning tasks:
	{\protect\circled{A}} We model the relationships among data columns, charts, and multiple-view visualizations using a three-layered heterogeneous graph;
	{\protect\circled{B}} For each node, the goal is to convert node features into a quality score;
	{\protect\circled{C}} Thus, the problem is abstracted into two sequence-to-one regression tasks at the chart- and MV-level, respectively.}
	\label{fig:graph}
\end{figure}

\name{} aims to assist in designing multiple-view visualizations (MV) for analyzing a dataset in tabular formats.
Since the design space of MVs is vast,
we consider the following constraints to make our research focused.
The design space of MV spans three dimensions: selection, presentation, and interaction among views~\cite{wang2000guidelines}.
We primarily focus on the selection of views,
\ie to select data columns for creating a chart,
and to select multiple charts for creating a cohesive whole.
That said,
other design factors such as layouts, colours, and interactivity are beyond our scope of automated recommendation.
Another simplifying assumption is that we support five primary chart types, including scatterplots, bar, line, pie, and area charts,
which are found most commonly used online~\cite{battle2018beagle}.
For visual encodings,
we only recommend visualization-level design choices (\ie chart types),
leaving encoding-level design choice (\eg color scales) to future work.
Besides,
we set the maximal number of encoded data columns per chart to be four,
as we find that 93.5\% charts in our training dataset encode no more than 4 data columns.
We make this decision to balance the trade-offs between resource usage and efficiency,
as the number of chart candidates grows exponentially with the cardinality of data columns.

It should be noted that we interchangeably use some terms in pair throughout the paper, including views and charts, as well as MVs and multiple-charts.
In the remaining text of this section,
we discuss the design considerations of \name{} and the overview of our method.

\subsection{Design Considerations}
The design of \name{} is guided by the following considerations:

\textbf{C1: Reducing the overheads of data selection.}
Given a data table,
data workers often need to select a few data columns for creating meaningful charts through trial and error.
This repetitive, tedious task should be delegated and automated.

\textbf{C2: Automating the selection for multiple charts.}
The system should select multiple charts from the candidate pool of exponentially growing size and further combine selected charts into a cohesive, meaningful multiple-view presentation.

\textbf{C3: Recommending charts conditioned on optional manual selections.}
Machine learning models are imperfect and could generate sub-optimal recommendations.
Besides, 
data workers often leverage their domain knowledge to specify partial selections regarding data columns or charts~\cite{wongsuphasawat2017voyager}.
Thus,
the system should blend human-input selections to provide conditional recommendations.

\textbf{C4: Leveraging the provenance data about authoring logs.}
The lack of MV corpora hinders the deployment of ML approaches~\cite{wang2020applying}.
We propose to leverage provenance as training datasets which receive growing research attentions through ``learning from users'' ~\cite{xu2020survey}.

\begin{figure*}[!t]
	\centering
	\includegraphics[width=1\linewidth]{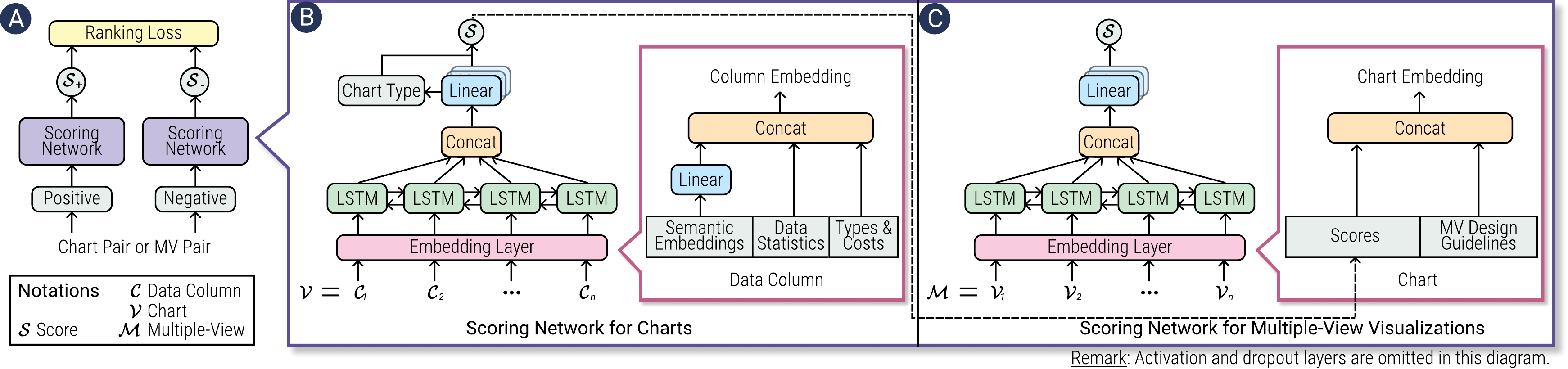}
	\caption{We propose a shared model architecture for both single-chart assessment and multiple-charts assessment:
	{\protect\circled{A}} We adopt a Siamese neural network structure which consists of two identical scoring networks working in tandem to compute comparable output;
	{\protect\circled{B}} The single-chart scoring network builds upon LSTM recurrent networks with data column embeddings mechanism, which includes semantic embeddings, data statistics, types, and other cost functions;
	and {\protect\circled{C}} The multiple-charts scoring network shares a similar architecture,
	and additionally leverages the output of single-chart scoring models and MV design guidelines to compute chart embeddings.}
	\label{fig:model}
\end{figure*}

\subsection{Method and Problem Abstraction}
\label{sec:overview:method}
\autoref{fig:overview} provides an overview of our system pipeline that consists of the automation, agency, and mixed-initiative module.
Given an input data table,
the automation module leverages deep learning models to predict an assessment score for all possible charts (\ie combinations of data columns) and MVs (\ie combinations of charts) (C1, C2).
Subsequently,
the scores are forwarded to recommend MVs through optimization,
\ie to generate an MV that maximizes the scores predicted by deep-learning models.
Critically,
the optimizer takes optional user specification as constraints to generate conditional recommendation (C3).
As users edit the recommended results,
their authoring logs are fed into the deep learning model in an offline manner (C4).

Our method highlights a perspective that treats the visualization construction process as a multi-layered heterogeneous graph.
As shown in~\autoref{fig:graph},
the graph consists of three layers where the nodes at each layer represent data columns, charts, and MVs, respectively.
Each link represents a ``composing'' relationship,
\eg several data columns compose a chart, and several charts compose an MV.
In this way, 
the node features are forwarded along the links to accomplish the machine learning task,
\ie to predict the assessment score.
For instance,
the chart feature is fused with the features with corresponding columns to predict the chart score.
Since the ``composing'' relationships can be expressed as a sequence (\eg an MV is a sequence of charts),
the score prediction problem is abstracted into a sequence-to-one regression task,
which is widely studied in deep learning research.

\section{Deep Learning Model}
\label{sec:model}
Our deep learning model aims to predict an assessment score for charts and multiple-view visualizations (MVs).
As discussed in \autoref{sec:overview:method},
we abstract this assessment problem into two sequence-to-one regression tasks at the chart- and MV-level, respectively.
This task abstraction allows us to design a shared architecture for both the chart assessment model and the MV assessment model.
In this section,
we first discuss the shared model architecture,
followed by descriptions of dedicated designs for single-chart assessment and multiple-chart assessment.

\begin{figure}[!ht]
	\centering
	\includegraphics[width=1\linewidth]{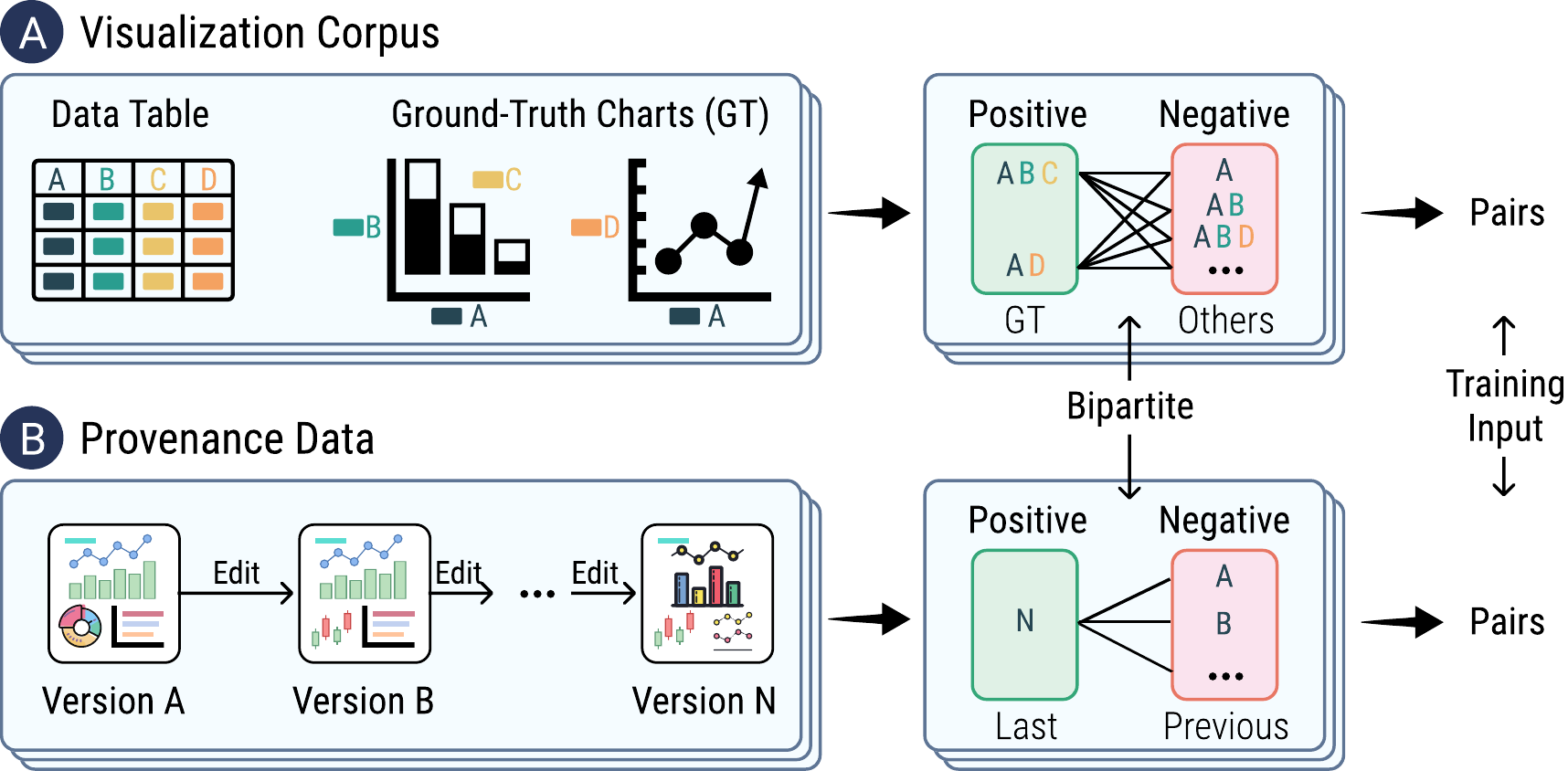}
	\caption{The training input contains chart pairs or MV pairs, which can be collected flexibly from {\protect\circled{A}} visualization corpus and {\protect\circled{B}} provenance data about authoring logs.
	The positive is considered better than the negative so that the model should learn to predict a higher assessment score for the positive than the negative.
	}
	\label{fig:dataCollection}
\end{figure}

\subsection{Model Architecture}
We propose to solve the assessment problem through a learning-to-rank approach~\cite{liu2011learning},
which aims to compute the overall rankings according to many partial orders.
Inspired by LQ2~\cite{wu2021learning},
we use a Siamese neural network structure~\cite{koch2015siamese} which consists of two identical scoring networks (\autoref{fig:model} \circled{A}).
Two scoring networks share the same weight and work in parallel to calculate comparable results.
Given an input ranked pair, denoted as $\langle \mathcal{I^+}, \mathcal{I^-} \rangle$,
where $\mathcal{I^+}$ is ranked higher than $\mathcal{I^-}$,
the scoring network, denoted $f$, is a function that converts an input to a score, denoted $\mathcal{S}=f(\mathcal{I})$.
Thus, 
the training goal becomes:
\begin{equation}
f(\mathcal{I^+}) > f(\mathcal{I^-}), \forall \langle \mathcal{I^+}, \mathcal{I^-} \rangle
\end{equation}

To that end,
we use the margin ranking loss, which imposes penalty to mistakes that assign a higher score to a lower ranked input:
\begin{equation}
\label{equ:pairLoss}
L(\mathcal{S^+}, \mathcal{S^-}) = max(0, m + \mathcal{S^+} - \mathcal{S^-}),
\end{equation}
where $m$ is a hyper-parameter for margins.

A key benefit of the above learning-to-rank problem formulation is the flexible and compatible mechanism for constructing and collecting training datasets,
that is, ranked pairs.
As shown in~\autoref{fig:dataCollection},
the ranked pairs can be collected from both visualization corpus and provenance data.
For visualization corpus (\autoref{fig:dataCollection} \circled{A}),
we consider the ground-truth charts to be better, \ie higher ranked, than all the remaining possible combinations of data columns.
For provenance data (\autoref{fig:dataCollection} \circled{B}),
we select the final output as the positive one, 
while the intermediate historical versions are marked negative.
We ignore the partial rankings among intermediate versions,
since the process of editing charts might be back-and-forth.
For example,
it is possible that users find the edited version to get worse.

Despite the flexibility and compatibility,
we acknowledge the downsides of this learning-to-rank problem formulation,
\ie the scalability.
Specifically,
we obtain pairs through bipartite matching between the positive and the negative.
However,
the number of possible charts and MVs grows exponentially with the cardinally of data columns and charts, respectively,
which theoretically results in an exponential time complexity.
Although we find the training time to be acceptable in our experiments,
future research should propose more efficient algorithms for selecting ``important'' pairs instead of enumeration to cope with the data explosion.

\subsection{Single-Chart Assessment}
We abstract the single-chart assessment into a sequence-to-one regression task,
which is widely studied in deep learning research for different types of sequences.
However,
this task in the context of visualizations poses distinct challenges from common sequences.
Common sequences such as natural language sentences and event sequences have a fixed vocabulary set,
\ie words and event types, respectively.
Correspondingly,
there exist many off-the-shelf methods for converting a vocabulary to a multiple-dimensional vector (\eg word embeddings),
which are often essential for deep learning models~\cite{mikolov2013efficient}.

In contrast,
most data columns in a data table tend to be distinct from each other and columns in different data tables.
In other words,
there does not exist a vocabulary set of data columns.
To solve this problem,
we adopt the strategy proposed by Zhou \ea~\cite{zhou2020table2charts,zhou2020table2analysis} to derive the data column embeddings through feature engineering.
Their column embedding mechanism consists of three types of information,
including the semantic text embeddings from column header text,
data statistics such as mean values,
and data types.
As shown in~\autoref{fig:model}~\circled{B}, 
the embeddings are fed into a bidirectional Long Short-Term Memory (LSTM) layer to learn the sequence features,
which are subsequently forwarded into linear layers to predict an assessment score regarding data column selections (denoted $\mathcal{S}(d)$ for a subset of data columns $d$).\begin{rev}
We implement an add-on feature for recommending visual encodings,
which is a basic component for practical visualization recommendation systems.\end{rev}
Specially,
we add another layer to predict the visual encoding (\ie five chart types including scatterplots, bar, line, pie, and area charts),
which is the likelihood (or posterior probability) of the data column selection belonging to chart type $v$,
denoted $\mathcal{P}(v)$.
Therefore,
the overall assessment score for both the data column selection and visual encodings is calculated by:
\begin{equation}
\label{equ:overallScore}
\mathcal{S}(d, v) = \mathcal{S}(d) \times \mathcal{P}(v)
\end{equation}

While we only consider the visualization-level encodings (\ie chart type) in this work,
other visual encodings can be added in a similar manner,
\ie to add a layer for predicting the visual encodings and compute the likelihood.
The key challenge here is that the detailed visual encodings vary among different chart types,
which requires simplifying abstraction (\eg VizML~\cite{hu2019vizml}) or comprehensive nested decision-tree-like models.
Besides,
it might need encoding visualization design rules to yield satisfactory performances (\eg Draco~\cite{moritz2018formalizing}).

There exists an alternative to our approach in~\autoref{equ:overallScore}.
To be specific,
we can train a model that directly learns $\mathcal{S}(d, v)$.
This can be achieved by adding the visual encodings to the training input, \ie the embeddings.
However,
a major downside is, again, the scalability,
as this alternative requires enumerating all possible combinations between data column selections and encodings.
This enumeration adds more degrees, depending on the cardinality of visual encodings, to the exponential complexity that ``explode'' the training time.

\subsection{Multiple-Chart Assessment}
\label{sec:model:multiple}
The multi-chart assessment model is similar to the single-chart assessment model,
since it shares a similar challenge that almost every chart is distinct.
However,
unlike charts,
there does not exist mechanism for computationally describing a multiple-view visualization (MV) in terms of data column selections and chart types.
Therefore,
we propose a novel method for modelling an MV as a sequence of chart embeddings.
\begin{rev}
The chart embeddings enable learning a shared MV representation,
\ie the MVs created for one particular dataset can be used as training data to recommend MVs for another dataset.
\end{rev}

\begin{figure}[!t]
	\centering
	\includegraphics[width=1\linewidth]{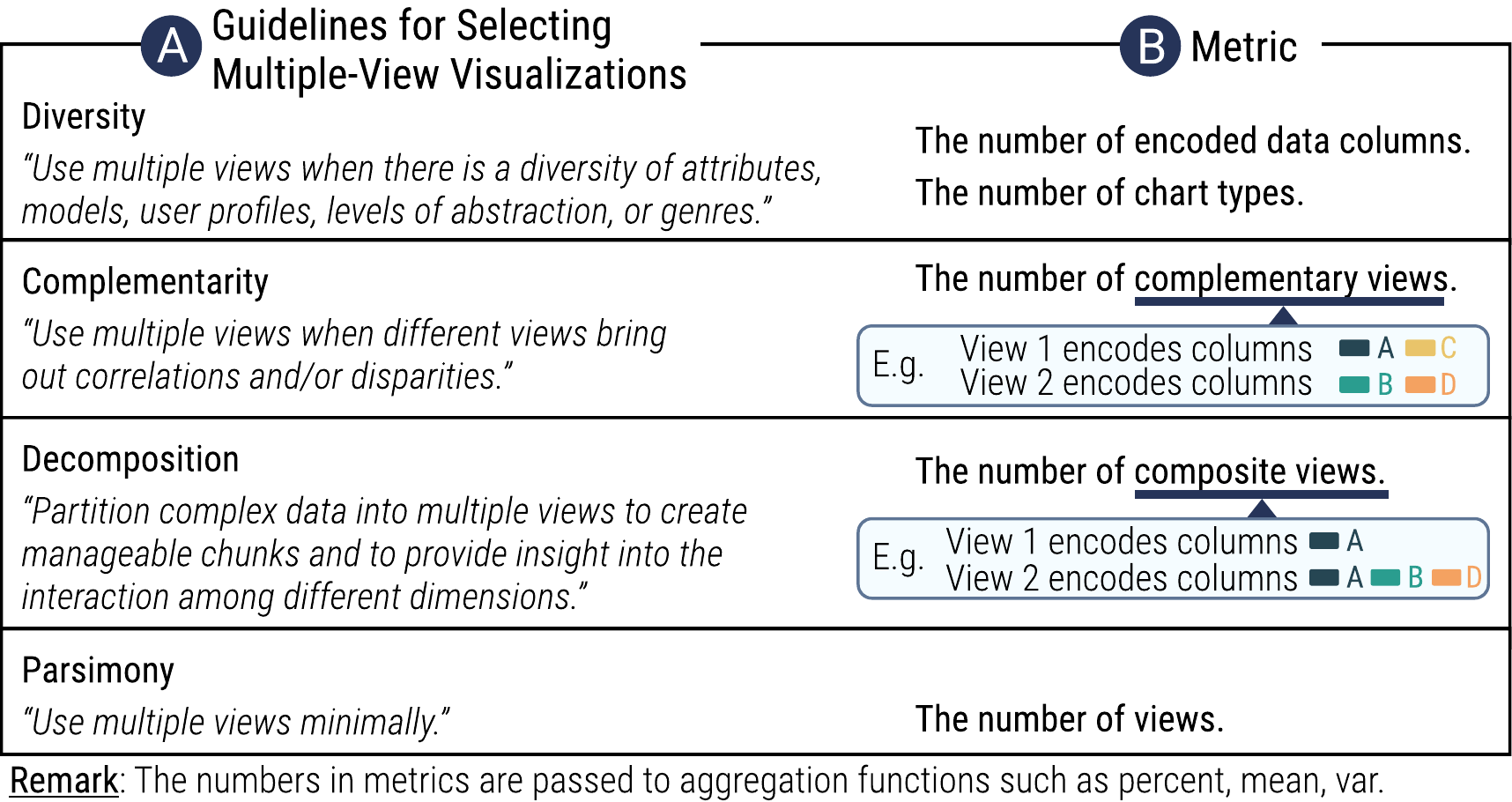}
	\caption{We encode {\protect\circled{A}} the guidelines for the selection of multiple-views in information visualization by Baldonado~\ea~\cite{wang2000guidelines} to {\protect\circled{B}}~computational metrics,
	which are subsequently fed into the deep learning models.
	}
	\label{fig:guideline}
\end{figure}

Our chart embedding mechanism involves two types of information.
First,
we use the scores predicted by the single-chart assessment model,
since our single-chart scoring network achieves satisfactory performances on assessment and encoding prediction.
Second, 
we propose to leverage design guidelines of MVs to improve the recommendation results,
inspired by Draco~\cite{moritz2018formalizing}.
As shown in~\autoref{fig:guideline},
we propose several novel functions to computationally encode the guidelines for the \textit{selection} of multiple views in information visualization proposed by Baldonado~\ea~\cite{wang2000guidelines}.
However,
the guidelines do not provide thresholds (\eg the maximal number of views) so that we cannot encode those guidelines to constraints in the same way with Draco.
Instead, 
we describe those guidelines as statistical features and fed it into deep learning models in an attempt to learn the ``best'' values of those guidelines (\autoref{fig:model}~\circled{C}).

It should be noted that Baldonado~\ea~\cite{wang2000guidelines} proposed another four guidelines for the \textit{presentation} and \textit{interaction} of MVs such as consistency and space optimization,
which are beyond the scope of this work and warrant future research.


\section{Interface}
\name~features an interactive interface that integrates the deep-learning-based recommendation into data exploration.
\autoref{fig:interface} shows the interface,
which is vertically divided into three panes.
The leftmost pane offers system-wise services including {\protect\circled{A}} uploading and viewing datasets,
{\protect\circled{B}} making MV recommendation,
and {\protect\circled{F}} other system functions such as changing themes.
The middle pane provides chart-wise functionalities including 
{\protect\circled{C}} editing charts and {\protect\circled{D}} browsing chart recommendation.
Finally,
users can explore and interact with the charts in the right-most pane ({\protect\circled{E}}).

\subsection{\mvRec}
\begin{figure}[!t]
	\centering
	\includegraphics[width=1\linewidth]{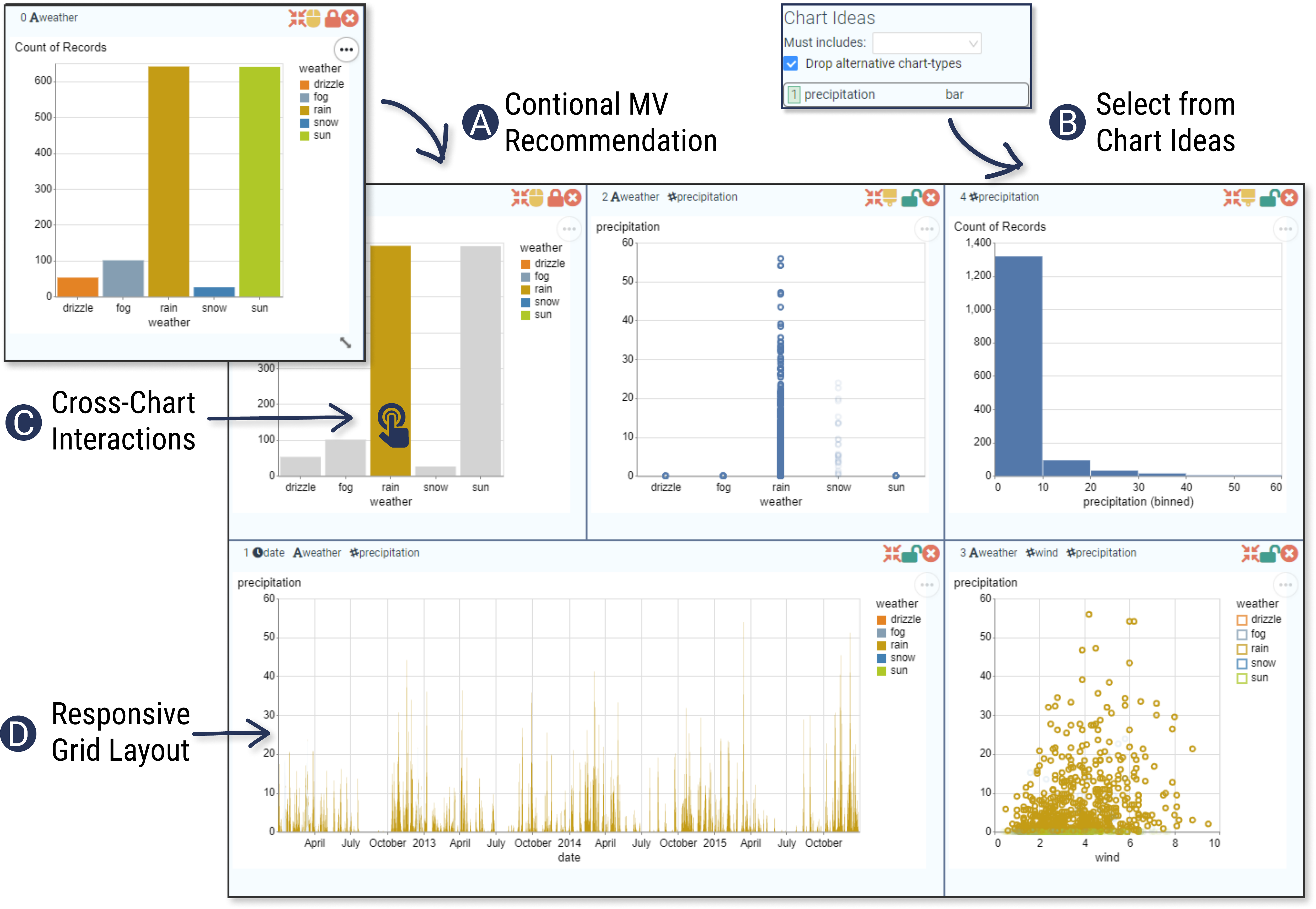}
	\caption{Major user interactions in \name:
	{\protect\circled{A}} Users request a  MV recommendation conditioned on the current selections of charts;
	{\protect\circled{B}} A chart can also be added from the \chartIdea;
	{\protect\circled{C}} Users can click or brush to perform cross-chart interactions;
	and {\protect\circled{D}} A chart can be moved, resized, and deleted in a responsive grid layout.}
	\label{fig:example}
\end{figure}

\mvRec~directly generates an MV based on user specifications.
To be specific,
it allows users to specify the number of charts in an MV and optionally locked charts.
Locked charts are taken into consideration during recommendation.
There is no limitation to the number of locked charts.
\autoref{fig:example} \circled{A} illustrates an example where \name~recommends a MV composed of five charts based on one user-locked chart.

We abstract the recommendation into a constrained optimization problem,
\ie to find a combination of charts that maximize the scores predicted by the multiple-chart assessment model (\autoref{sec:model:multiple}).
However,
this is a challenging problem due to the huge, exponentially growing space of chart combinations.
For the sake of running time and user experience,
we implement a naive greedy algorithm that takes locked charts as the initial selection and incrementally picks one chart that leads to the highest score.
In this way,
\name~can make a recommendation responsively within few seconds,
which, however, might not always yield the optimal solution.
Future work should study and propose advanced algorithms to further improve the performance,
\eg possibility-based approximation algorithms~\cite{luo2018deepeye}.

\subsection{\chartIdea}
\chartIdea~offers an alternative method for users to interact with automatic recommendations.
As shown in~\autoref{fig:interface}~\circled{D},
it provides chart recommendations based on the current MV.
Specifically,
it lists and ranks charts from top to bottom in the order of the assessment score,
\ie the score of the MV after adding the chart to the current MV.
This process shares the same computation procedure with \mvRec{} since it can be seen as one step in the greedy algorithm.
User could click a chart to add it to \mvView~(\autoref{fig:example}~\circled{B}).
To facilitate browsing and selecting charts of interest,
users could select data columns in the ``Must Includes'' drop-down list to view charts containing selected columns.
\mvView~by default considers charts encoding the same data columns to be the same, irrespective of the chart type.
Users could un-check the ``drop alternative chart-types'' to view all alternative chart types.

Different from \mvRec~which is passive and triggered upon request,
\chartIdea~makes active recommendation,
\ie it is automatically updated upon user changes to the current MV,
including adding, editing, and removing charts.
We design those two recommendation strategies in an attempt to probe into the open challenge in designing mixed-initiative user interfaces,
\ie ``to what degree will such systems promote behavior characterized by user control vs. more passive acceptance of algorithmic recommendations?''~\cite{heer2019agency}.
We reflect on our findings about the active and passive recommendation in~\autoref{userStudy:discuss:agency}.

\subsection{\chartEd}
\chartEd~allows users to edit chart specifications,
including visual encodings, chart types, and data transformation.
We implemented \chartEd~since users might need to make adjustments to recommended results or create charts that satisfy personal needs.
We borrowed the editor design in Voyager~\cite{wongsuphasawat2015voyager} which has proven to be easy to use.
To be specific,
\chartEd~supports editing chart specifications in the Vega-Lite grammar~\cite{satyanarayan2016vega}.
For example, \autoref{fig:interface}~\circled{C} illustrates the visual specification of the top-right chart in \autoref{fig:interface}~\circled{E},
which is highlighted with the same background color.
\name~currently supports five chart types including area, bar, line, pie charts and scatterplots that are found most commonly used online~\cite{battle2018beagle}.

Since the deep learning model only recommends chart types,
we decided the visual encodings according to the heuristics in Voyager~\cite{wongsuphasawat2017voyager},
\eg nominal data types are firstly assigned to the x/y channel.
\chartEd~provides six encoding channels, including x, y, column, row, size, and color.
We acknowledge that those heuristics might return sub-optimal visual encodings.
Future work could implement advanced visual encoding recommender such as Draco~\cite{moritz2018formalizing} and InfoColorizer~\cite{yuan2021infocolorizer} to improve the system.

\subsection{Presentation and Interaction}
\label{sec:interface:log}
Although \name~primarily focuses on the selection of MVs,
we provided basic support for interaction and presentation of MVs.
As shown in~\autoref{fig:example}~\circled{C},
users could perform cross-chart filtering between multiple charts through mouse clicking or brushing.
Besides,
users could move and resize a chart in a responsive grid layout (\autoref{fig:example}~\circled{D}).
Finally,
users could change the theme, restore historical versions, and save logs in~\controlP ~(\autoref{fig:interface}~\circled{F}).
The logs contain the uploaded data table, 
the computed features of the data table, 
event logs for the functionality above,
as well as the rendered charts of all historical versions.
It is up to users to decide whether the logs will be stored.

\section{Evaluation}
To evaluate \name,
we conduct an experiment to evaluate the model performance from the perspective of techniques,
as well as a user study to understand \name~from a system perspective.

\subsection{Model Experiment}
We evaluate the algorithm performance of our deep learning model as introduced in~\autoref{sec:model}.
We report the dataset, baselines, implementation details, results and discussions in the following text.

\subsubsection{Dataset}
We collect and construct two datasets for single-chart and multiple-charts assessment, respectively.
For single-chart assessment,
we adopt the Excel corpus in Table2Charts~\cite{zhou2020table2analysis} which consists of 271,250 charts from 167,329 data tables.
After filtering out data tables consisting of more than ten columns (14.2\%) and charts encoding over four columns (6.5\%),
we obtain 3,920,941 pairs following the procedure described in~\autoref{fig:dataCollection}~\circled{A}.
Since the number of possible chart pair is huge,
we only keep pairs where two charts encoding the same number of data columns.
Each chart in a pair is described as a sequence of data columns with feature vectors sized $4 \times 96$, where $4$ is the maximal sequence length and $96$ is the feature size per column.

However, 
the aforementioned Excel dataset is not suited to multiple-chart assessment,
since most Excel sheet consists of a single chart (75.6\%). 
Thus, 
we decide to use the user logs of editing MVs as the training dataset, as illustrated in~\autoref{fig:dataCollection}~\circled{B}.
Here we report the performance over the provenance data that we obtained from the user study (\autoref{sec:eva:userStudy}),
which consists of 692 pairs from 11 participants\footnote{One participant decided not to share the logs with us.}.
Each chart is described as a feature vector sized $9 \times 1$,
and the maximal sequence length of an MV is 12 in our dataset.

\subsubsection{Comparison with Baseline Model Structures}
Since existing machine-learning-based visualization recommenders do not focus on generating multiple charts,
we do not identify a fair baseline recommender for comparing the recommendation performances.
Instead,
we compare our model structure in~\autoref{fig:model} against two machine learning model structures used in previous visualization recommenders:

\begin{compactitem}
    \item \textbf{NN} (Neural Network) is the best-performed method in VizML~\cite{hu2019vizml} and LQ2~\cite{wu2021learning}, which consists of several fully-connected layers and ReLU activation layers.
    \item \textbf{RankSVM} (Ranking Support Vector Machine)~\cite{lee2014large} is adopted in DeepEye~\cite{luo2018deepeye} and Draco~\cite{moritz2018formalizing}. 
    Given a ranked pair $(v_1, v_2)$ with feature vectors $x_1, x_2$, it predicts the order by $\sign f(x_1 - x_2)$, where $\sign$ is a sign function and $f$ is the training target.
\end{compactitem}

\subsubsection{Implementation and Model Detail}
We implement our deep-learning models using PyTorch~\cite{paszke2019pytorch}.
\begin{rev}
We manually tune the hyper-parameters of our and baseline NN models by diagnosing the training curves,
\ie until both the validation and training scores converge to a desired level with small gaps in between them. 
We find that the margin hyper-parameter in the loss function (\autoref{equ:pairLoss}) plays a dominant role in the performance,
while other parameters such as dropout rates and learning rates have a smaller effect on the model behaviour.
\end{rev}

We run the experiment on a Linux server with a Tesla P100 GPU.
The number of epochs is set to 10.
We split the dataset with an 80/20 training-testing ratio and run Monte-Carlo cross validation~\cite{xu2001monte} 10 times.

\begin{table}[!t]
\caption{Model performance in terms of the ranking accuracy on pairs (\%) through Monte-Carlo Cross-Validation averaged for 10 runs with an 80-20 training-testing split ratio.}
\label{table:results}
\centering
\begin{tabular}{lccc}
\toprule
                  & \textbf{Ours} & \textbf{NN} & \textbf{RankSVM} \\ \hline
\textbf{Single Chart}  &  \underline{97.86}    &  96.59     &  93.42  \\
\textbf{Multiple Charts} &  \underline{94.05}    &   88.99      & 84.78  \\ \bottomrule
\end{tabular}
\end{table}

\begin{figure}[!b]
	\centering
	\includegraphics[width=1\linewidth]{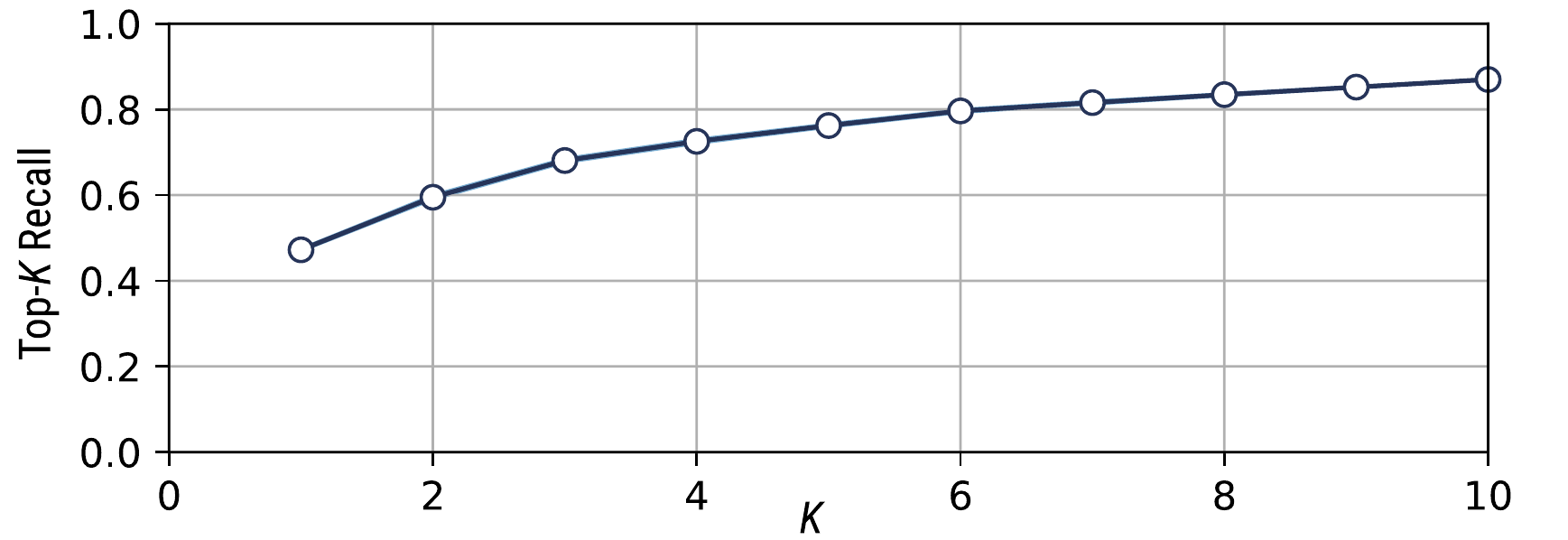}
	\caption{The top-$k$ recall curve of our model averaged after Monte-Carlo Cross-Validation. Recall at $k$ is the proportion of ground-truths found in the top-$k$ recommendations.}
	\label{fig:topKRecall}
\end{figure}

\subsubsection{Result and Discussion}
\autoref{table:results} presents the performance of our models and baselines.
Our model outperforms baseline approaches in both single-chart and multiple-chart assessment tasks.
We note that multiple-chart assessment performs poorer than single-chart assessment, suggesting that our feature engineering methods for describing an MV might be insufficient and unrepresentative.
Therefore,
future research should propose efficient methods for modelling and describing MVs.

The results in~\autoref{table:results} should be interpreted carefully since it represents the ranking accuracy on the pairs.
In other words,
it does not directly reflect the quality of recommended results.
To better understand the performance of recommendation,
we run an additional experiment to evaluate the top-$k$ recall on single-chart assessment,
\ie the proportion of ground-truth charts found in the top-$k$ recommendations.
As shown in~\autoref{fig:topKRecall}, our model on average reaches 47.2\%, 68.1\%, 76.2\%, and 87.0\% at $k=1,3,5,10$, respectively.
That said,
87\% of the ground-truths can be found within the top-10 recommended results.
The standard deviation is small, \ie $< 0.01$, suggesting the model stability.
It should be noted that we do not run this additional experiment on multiple-chart assessment at the current state,
since the data size is small ($N = 11$) and thus insufficient in proving statistically meaningful results.
Moving forward,
we are excited to involve more participants, collect large-scale MV datasets, and further investigate the performance.

Critically,
our model achieves satisfying performance at the cost of algorithm complexity and training time.
Our method for collecting training data (\autoref{fig:dataCollection}) can be considered a ``brute-force'' strategy that enumerates possible pairs between good and bad selections.
This brute-force manner is expensive,
\eg we convert 271k charts into 3,920k pairs. 
Subsequently, this growth imposes additional costs on the computational resources for deep learning.
Thus, continued research on problem formulation and model architecture will be beneficial to the resource-performance tradeoff.


\subsection{User Study}
\label{sec:eva:userStudy}
We conducted a formal user study to evaluate \name~from a system perspective.
Our overall goal is to investigate: (1) whether \name~helps data workers create multiple-view visualizations and conduct data analysis; 
and (2) whether the deep-learning-based recommendation assists and engages users in exploring tabular datasets.

\subsubsection{Study Design}
Based on the above goals, 
we conducted a usability test to understand how users interacted with \name~and gather their feedback.

\textbf{Participants.}
We recruited 12 participants including five females.
We asked participants to specify their experience in charting software (\eg Excel, Tableau) and programming tools (\eg Matplotlib, Vega) on a five-point Likert scale, 
where 1-point denotes ``never heard of it'' and 5-point means ``very familiar''.
Their responses suggested varying degrees of experience in data visualizations,
\ie charting software ($\mu = 3.75, \sigma=0.75$) and programming tools ($\mu = 2.92, \sigma=1.51$).
\begin{rev}
Two participants (E1,2) were experts in data analysis and visualization from a global high tech company.
Remaining participants (P1-10) were undergraduate or postgraduate students with different majors including computer science, mathematics, and finance.
\end{rev}

\textbf{Datasets and Settings.}
We prepared five datasets from Vega-lite dataset\footnote{https://github.com/vega/vega-datasets},
including cars, gapminder, penguins, countries, and Seattle-weather.
Those datasets consisted of 6-9 data columns,
which had a suitable level of complexity and exploration efforts for running user studies.
\begin{rev}
To train the initial multiple-chart assessment model and enable recommenders,
we conducted a pilot study to collect a small dataset of editing logs.
We only used the cars dataset in the pilot study to qualitatively investigate how well the trained models generalize to other new datasets.
\end{rev}

\textbf{Procedure.}
A study session lasted between 50-60 minutes.
We first asked participants to read and sign the consent form,
especially that ``I understand that my editing logs and comments will be collected''.
Participants were then given a tutorial about \name~and allowed to try out and get familar with the system ($\sim$15 minutes).
Next, we instructed participants to ``select a dataset that you have not heard of and conduct data analysis'' ($\sim$20 minutes).
For the sake of accountability,
we asked participants to report findings during exploration and ``finally create an MV to communicate important findings''.
The study ended with an exit questionnaire on the subjective ratings of \name{} functions and a short interview.

\begin{figure}[!t]
	\centering
	\includegraphics[width=1\linewidth]{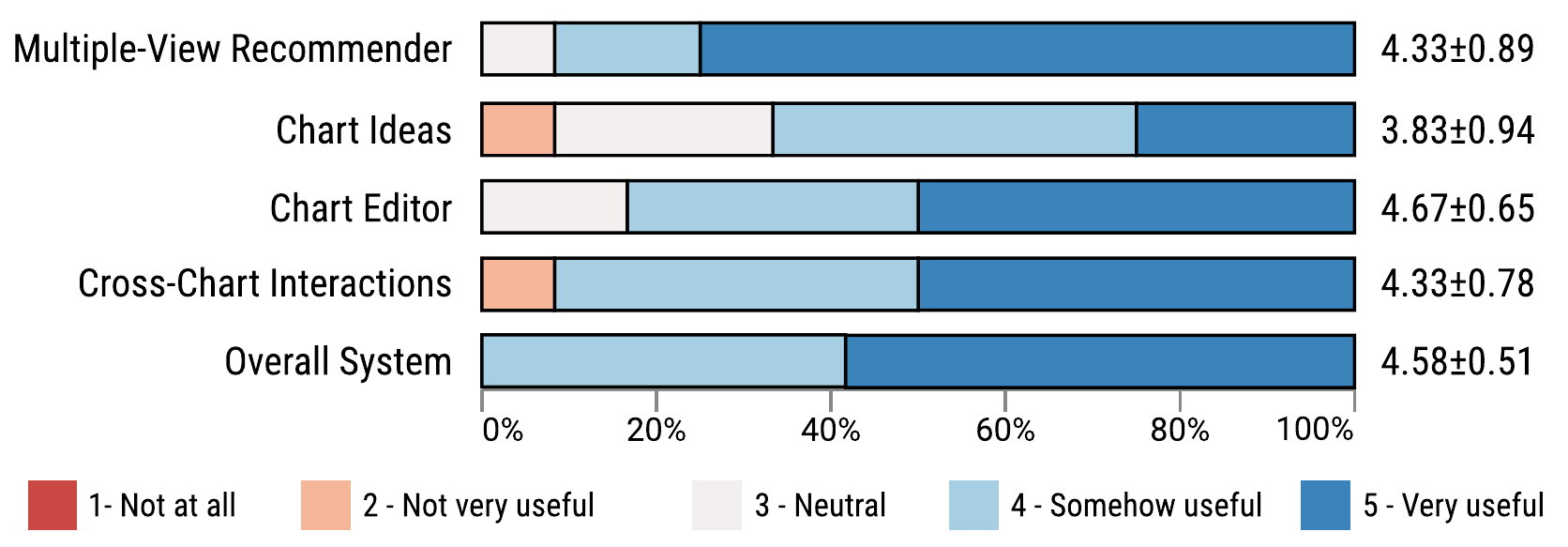}
	\caption{Usefulness ratings for 4 features and the overall system on a 5-point likert scale ($N=12$). The rightmost column indicates the average and standard deviations.}
	\label{fig:questionnaire}
\end{figure}

\textbf{Interview Questions.}
We asked participants' opinions on the pros and cons of each functionality in \name~and the overall system.
Then we asked them to compare \name~with their previous experience in analyzing tabular datasets, and if any, in multiple-view visualizations or dashboards.
To elicit fair comments,
questions were one-off to avoid being double-barreled,
and we avoided potentially leading questions such as ``how about \name~in terms of conveniency''~\cite{litwak1956classification}.
As such,
we collected explicit immediate reactions from participants.
Although this setting might have hindered us from assembling in-depth thoughts through detailed inquiries,
our expectation was that immediate responses reflected the foremost thoughts about \name.
The interview session ended with questions regarding the performance of automated recommendation and the user experience about the mixed-initiative system.

\subsubsection{Participant Feedback}
Overall, \name~received an enthusiastic reaction from the participants.
As shown in~\autoref{fig:questionnaire},
participants, in general, deemed the functionality of \name~and the overall system to be useful.
In the following text,
we summarized participants' feedback on the advantages of~\name.

\textbf{\name~is useful, convenient, and easy to learn.}
From a tool perspective,
\name~scored a high usefulness rating among participants ($\mu = 4.58, \sigma=0.51$).
When asked to compare \name~with their previous experience with charting software or libraries for analyzing tabular datasets,
eleven participants explicitly commented on the convenience of~\name,
\ie it required few efforts to browse, select, create, and edit multiple charts.
Four participants (P5,6,9,10) emphasized that \name~has a short learning curve, 
\eg ``it is a plain and simple tool for the masses''.

\begin{figure}[!t]
	\centering
	\includegraphics[width=1\linewidth]{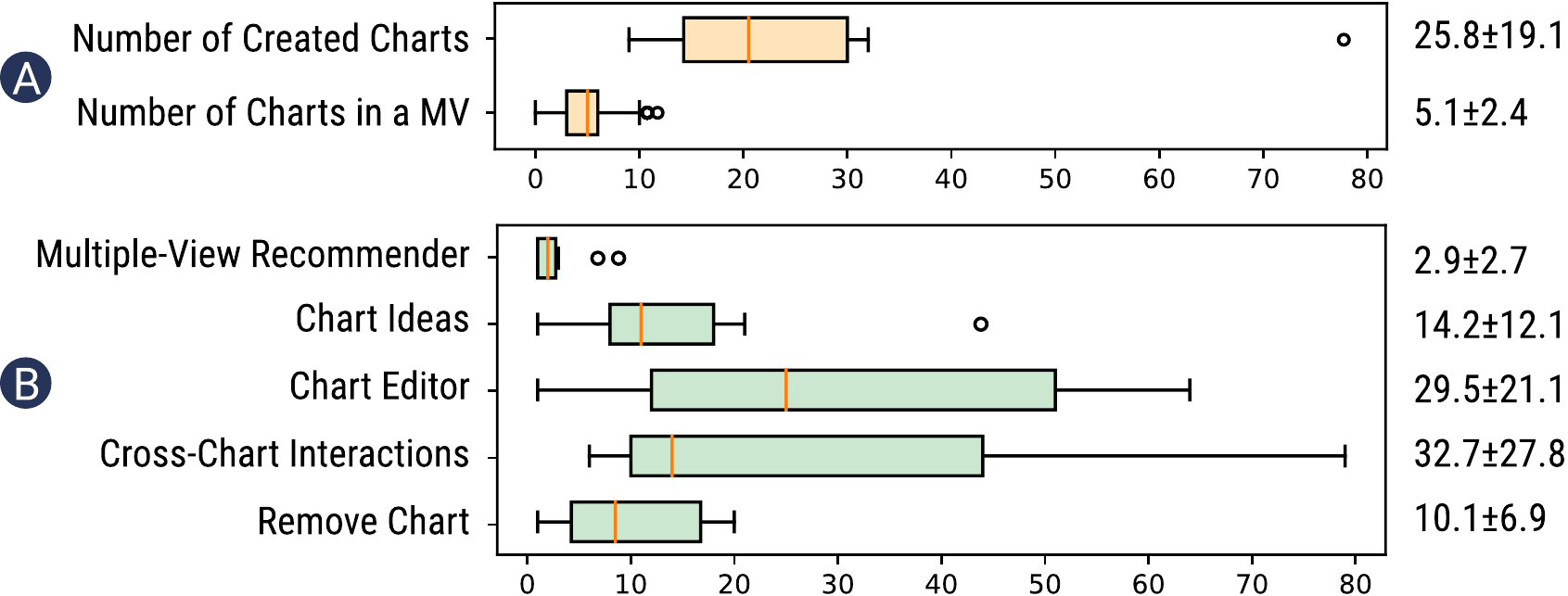}
	\caption{Analysis of user behaviour in the user study: 
	{\protect\circled{A}} The chart usage; and 
	{\protect\circled{B}} The usage count of each functionality.}
	\label{fig:logs}
\end{figure}

\textbf{\name~facilitates exploratory data analysis.}
Participants expressed that the recommendation provides helpful starting points (P5,8,10) or inspiration when they get stuck (P9).
E2 positioned the prototype to exploratory data analytical scenarios when users had little understanding and no clear analytic tasks about the dataset.
Nine of twelve participants made their first attempt at creating multiple-view visualizations (MVs),
all appreciating that MVs helped them explore datasets from different perspectives effectively.
P4 who had previously created dashboards for the purposes of monitoring commented that ``it takes some yet small costs to learn (the use of MVs in \name). Everything is intuitive''.

\textbf{\name~weighs automation and human intervention.}
Participants exhibit generally satisfactory yet diverse feelings on the ``overall performance of machine learning models'' during interviews.
Their responses from ``roughly 40\% of results need manual adjustments'' (P10) to ``I tried to create charts myself but found the recommended results to be better than my decisions'' (P3).
This was further evidenced by the large negative correlation ($\rho = -0.42$) between the usefulness ratings of \mvRec{} (automation) and \chartEd{} (human intervention),
\ie the more participants consider automation to be useful, the less they appreciate manual efforts.
During interviews, all participants appreciated the ``semi-automated'' or ``human-computer collaborative'' paradigm of \name,
\ie users could modify the automatically recommended charts which are dynamically updated in line with their current selections.

\subsubsection{Log Analysis}
We investigated the user behaviour about the usage of charts and each functionality,
which is described in~\autoref{sec:interface:log}.
As shown in~\autoref{fig:logs},
participants on average created 25.8 charts which indicated a high user engagement.
The number of charts in an MV during exploration averaged 5.1,
showing that overmuch charts were not desirable.

We observed differences among the usage count of each functionality.
Participants had engaged more in \crossChart{} and \chartEd{},
showing that participants had actively interacted with automated support and engaged in their own creative practice.
Those high usage counts also conformed to the high usefulness ratings of those two functionalities.
From the aspect of recommendation,
\chartIdea{} was much more frequently used than \mvRec{},
which was in contrast to the usefulness ratings.
During interviews,
participants explained that they used \mvRec{} at the beginning and found it helpful to offer initial ideas.
Afterwards,
they preferred to use \chartIdea{} to create and explore charts incrementally.
However, 
\chartIdea{} suffered few minor design flaws such as lacking previews that lowered their rankings.

\section{Implication and Lessons Learned}
In this section,
we reflect on the lessons learned for designing more effective MVs, balancing agency and automation, and creating visualization authoring tools.

\subsection{Designing More Efficient MVs}
All participants, including those who were new to MVs,
considered MVs to be useful and intuitive for data analysis.
\begin{rev}
Despite enthusiastic feedback,
we identify several future directions to further improve the usefulness and user experience (UX) of MVs.
\end{rev}

\textbf{Integrating methodology of visual analytics.}
E2 drew on the high autonomy and freedom in \name,
saying that our design was proper in terms of user experience but might lack guidance for laypersons.
This comment was exemplified by P7, a beginner to MVs who commented ``I had no idea how to analyze data with MVs at the beginning''.
Different from P7,
some other MV beginners explained their methodology,
\eg ``I want to see the distributions of each variable first and then explore the correlations'' (P1),
to which we refer as breadth-first exploration.
In contrast,
other participants (P2,8,9) adopt depth-first exploration,
\eg ``I focus on (a variable) and examine how it influences others''.
Those comments underscore the importance of methodology for efficient data analysis with MVs.
Therefore,
future research should better understand the analytical pipelines of MVs and integrate them into automated creation of MVs to provide guided exploration,
\eg ``overview-first, detailed on demaind''.

\textbf{Enhancing the interactions among views.}
Participants thought highly of the usefulness of \crossChart{} ($\mu = 4.33, \sigma=0.78$),
commenting that interactions helped reveal new insights (P6,8),
scope to data of interest (P7), and reduce clutter (P3,10).
A major UX issue arises that ``it is unclear which charts are connected'' (P1).
P1 suggested to put linked charts nearby,
and P2 recommended explicit representations of inter-chart relationships such as an addition figure showing the relationships.
Besides,
P11 noted that ``clicking and brushing is insufficient''.
Those comments call for continued research on the presentation and interaction of MVs.

\textbf{Modeling the analytical progress of MVs.}
\begin{rev}
\name~implements a linear model for tracking and restoring authoring history,
which is common in UX design such as text editors.
However, we find that this linear model is insufficient in visualization authoring.
Specifically,
participants often experienced iterative and back-and-forth exploration processes,
\ie to reach back previous charts and explore alternative chart combinations.\end{rev}
Future research might represent analytical progress using non-linear, tree-like models that better reflect the iterative changes of MVs.
Besides,
they expressed demands for more detailed history tracking due to the huge pool of charts,
\eg ``there are many charts. I want the system to tell which charts I have checked'' (P6) and ``I need to know columns that I have not examined'' (P8).

\textbf{Recommending theme-based and explainable MVs.}
Although all participants agreed that MVs promoted data analysis,
some participants pointed that a single MV might be insufficient.
For instance,
P8 wished to create multiple MVs, each focusing on several data columns.
Similarly,
E2 commented that a dashboard typically consists of 4-6 views to avoid being overwhelming and embody a core theme.
Besides,
participants felt that it was not always easy to understand the recommended MV (P1,2,6) and a ``readme'' might help (P4).
Therefore,
an interesting question is how to summarize an MV into a theme,
which in turn could inform research in recommending more explainable theme-based MVs.
Besides,
it is helpful to conduct user studies with laypersons (\eg~\cite{shu2020makes}) to investigate the understandability of MVs.

\subsection{Balancing Agency and Automation}
\label{userStudy:discuss:agency}
E2 describes \name~as a ``conversation'' between users and machines - both could update the current MV according to the feedback from another.
In this subsection, we reflect on the participants' feedback about their ``conversations'' with automation.

\textbf{Automation might be disrupting.}
\name~offers two methods for recommending MVs, including \mvRec{} that generates multiple charts at once and \chartIdea{} that recommends a single chart dynamically upon each manual change to the MV.
\chartIdea{} received a lower rating of usefulness ($\mu = 3.83, \sigma=0.94$) than the former ($\mu = 4.33, \sigma=0.89$).
Although participants generally agreed that it was helpful to make recommendation bases on the current MV,
concerns arose that each update required rethinking the newly recommended charts and thus sometimes became disturbing (P5,7,8 and E1,2).
Besides, the rethink was cognitively costly as \chartIdea{} listed charts as text descriptions which were less intuitive than chart images.
Future research should investigate how to alleviate this problem to leverage the combined power of agent and automation.

\textbf{Personalization is in-demand.}
Participants were inconsistent in conducting data analysis with \name,
\eg breath-first and depth-first exploration.
As such,
they exhibited conflicting feelings over the recommended results, 
including ``lacking diversity'' (P6,11) and ``diverging from my current selection'' (P8).
Similar opinions include that ``it did not learn that I dislike pie charts although I had removed pie charts for many times'' (P3).
Those problems require further research on recommendation systems that counter-balance the short-time user satisfaction and long-term model converge~\cite{schnabel2018short}.
One promising solution might be personalized recommendations that allow users to specify their intent on different criteria (\eg Calliope~\cite{shi2020calliope}).
Besides,
future recommendation system should propose collaborative filtering approaches~\cite{su2009survey} by grouping similar peer users and generating recommendations using the neighbourhood.

\textbf{Leveraging user data warrants deeper studies.}
\begin{rev}
We proposed to recommend MVs by learning from user provenance data.
Despite being useful and generalizable,
such systems suffered from the ``cold-start'' problem that the recommended results were far from being perfect due to limited training datasets at the beginning.
Thus,
a clear next step is to collect more provenance data on new datasets to improve the performance and better evaluate the system.
Another interesting problem is to explore strategies such as few-shot learning to cope with the ``cold-start'' problem.\end{rev}
When asked for opinions for gathering user data,
all participants exhibited open-minded attitudes if ``I am informed''.
However, concerns regarding the data quality arose among participants,
such as ``I am not confident that my data is of high-quality and helpful'' (P5) and ``many of my operations were meaningless'' (P9).
E2 similarly expressed this point, saying ``charts are similar to paintings. It is easy to differentiate `rubbish' but much harder to compare `masterpieces'. How to ensure the quality of collected logs?''.
Therefore,
it warrants more studies to understand how to leverage user data effectively (\eg data wrangling and cleaning).

\subsection{Creating More Useful Charting Tools}
From a charting tool perspective,
\name~receives enthusiastic comments on its convenience, usefulness, and short learning curves.
Still, participants required more functionality for further improvements. 

\textbf{Data Transformation.}
Nine participants expressed the needs for applying data transformation to the raw data table.
\name~supported limited operations of data transformation (\eg bin) that were insufficient.
A promising research question would be to predict the data transformation given a data table.
Besides,
E1 suggested leveraging natural language interface for specifying data transformation,
which was easy-to-use for novice users and could be integrated into \name. 

\textbf{Expressiveness.}
In general, participants felt that the chart types supported in \name~were sufficient in basic data analysis and rated highly of the Chart Editor ($\mu = 4.66, \sigma=0.65$).
This feedback conformed to our findings in the training dataset that those basic types were dominant.
However, 
two participants demanded additional chart types such as box-plots (P2,11).
In addition,
participants presented various requirements for editing access to scales (P4,10) and text (E2).

\section{Conclusion and Future Work}
\begin{rev}
We present \name, a mixed-initiative system that leverages deep-learning-based recommendation for creating and authoring dashboards to analyze a data table.
We believe that this is an important and challenging problem and there are several open directions for future work.\end{rev}

\textbf{Investigating visualization-tailed machine learning.}
From a broader sense,
it remains an open challenge to propose proper machine learning model and feature representation techniques for visualization research~\cite{wu2021survey,wang2020applying,chen2019towards,yuan2021deep}.
Future research should investigate and propose advanced deep learning models,
\eg generative models~\cite{wang2019deepdrawing}.
\begin{rev}
Besides,
we see that Graph Neural Network (GNN)~\cite{scarselli2008graph} is promising,
which has recently demonstrated success across various application domains.
For instance, 
Graphiti~\cite{srinivasan2017graphiti} models tabular datasets as a homogeneous graph where a node represents a column and a link indicate the linking conditions.
As shown in~\autoref{fig:graph},
we could extend this model to a heterogeneous graph where data columns, charts, and MVs are nodes and an edge represents a composing relationship (\eg several columns compose a chart).
In this way,
the recommendation problem could be abstracted into edge regression or edge cutting tasks that might be solved by GNNs.\end{rev}
However,
the graph model of MVs suffer an ``open-vocabulary'' problem,
\ie each node, including columns, charts, and MVs, is distinct.
This differs from the application of GNN in other domains where there exists a fixed, closed vocabulary of nodes (\eg scholars in citation network).
Future research should be aware of the particularity of visualization research and study visualization-tailored solutions.

\textbf{Continuing research on multiple-view visualizations.}
Despite the wide usage of multiple-view visualizations,
research efforts on MVs remain relatively limited.
From a theoretical perspective,
there exist limited understandings about the guidelines and the design space of MVs~\cite{chen2020composition}.
Thus,
it is vital to continue research on the empirical studies of MVs to understand the user engagement, the strategies for data analytics, design considerations and ``best practices'' of MVs.
From a practical perspective,
there are limited grammars or language that allow describing MVs in a shared representation.
Shared representations are critical since they can be authored and edited by both human and machines,
which in turn contribute to or learn from solutions to design better visualizations~\cite{heer2019agency}.
Recent research (\eg~\cite{hu2019vizml,wu2020mobilevisfixer}) has suggested that declarative grammars (\ie parameters) are effective and compact representations for data visualizations that benefit the use of machine learning,
while visualization images are expensive and inefficient representations~\cite{wu2021survey,haehn2018evaluating,fu2019visualization}.
That said,
continued research on declarative grammars for the presentation and interactions of MVs is essential.

\textbf{Benchmarking recommendation and systems.}
In both model experiment and user study,
we do not directly compare \name~with a baseline approach.
This is because we do not identify a fair baseline for both the deep learning model and the system.
Specifically,
existing visualization recommenders do not focus on generating an MV.
\begin{rev}
Sequential single-chart recommenders do not consider the relationships among charts (\eg DeepEye~\cite{luo2018deepeye}) or target at a logically coherent data story for storytelling purposes (\eg Calliope~\cite{shi2020calliope}) instead of a cohesive, linked MV for analyzing different perspective of data.\end{rev}
Thus, our experiments compare our deep-learning model with the machine learning methods in existing recommenders on our dataset.
In the future,
we plan to compare models using other datasets (\eg DeepEye~\cite{luo2018deepeye}) to better evaluate our approach.
Similarly,
we do not find a fair baseline system that supports both editing multiple-view visualizations and providing recommendations.
We hope that our initial results will inspire and provide benchmarks for future work.

\textbf{Understanding users in mixed-initiative visualization systems.}
\name~only takes one of the initial steps in integrating deep learning models into visualization authoring systems.
We draw on participants' feedback to discuss the effects of different recommendation strategies.
We found that active recommendation might be disrupting while passive recommendation are much less likely to be used. 
That said, it remains challenging to propose approximate interaction designs that prompt efficient usage and understandings of automation while reducing disruptions to users.
Another exciting idea is to characterize user behaviour~\cite{chen2019lassonet} and make recommendations through collaborative filtering approaches~\cite{oppermann2020vizcommender}.
Finally,
it is important to consider visualization tools as a social system where users could share and communicate their visualizations with each other.
Recent research in mining user behaviour in visualization systems~\cite{monadjemi2020competing} seems a promising direction.



\acknowledgments{
The authors wish to thank anonymous reviewers for their suggestions.
This work was supported in part by
a grant from XYZ.
}

\bibliographystyle{abbrv-doi}

\bibliography{main}
\end{document}